# Truly concomitant and independently expressed short- and long-term plasticity in Bi$_2$O$_2$Se-based three-terminal memristor

Ziyang Zhang[1+], Tianran Li[2+], Yujie Wu[1+], Yinjun Jia[3+], Congwei Tan[2], Xintong Xu[4], Guanrui Wang[1], Juan Lv[1], Wei Zhang[3*], Yuhan He[5], Luping Shi[1*], Hailin Peng[2*], Huanglong Li[1*]

[1]Department of Precision Instrument, Center for Brain Inspired Computing Research, Tsinghua University, Beijing, 100084, China [2]Center for Nanochemistry, Beijing Science and Engineering Center for Nanocarbons, Beijing National Laboratory for Molecular Sciences, College of Chemistry and Molecular Engineering, Peking University, Beijing 100871, China School of Life Sciences, Tsinghua-Peking Joint Center for Life Sciences, IDG/McGovern Institute for Brain Research, Tsinghua University, Beijing 100084, China [4]School of Aerospace Engineering, Tsinghua University, Beijing, 100084, China [5]Department of Electronic Engineering, Tsinghua University, Beijing, 100084, China

[+]These authors contributed equally

Emails: zhangweilab@biomed.tsinghua.edu.cn
lpshi@mail.tsinghua.edu.cn
hlpeng@pku.edu.cn
li_huanglong@mail.tsinghua.edu.cn

**Abstract**

Orchestration of diverse synaptic plasticity mechanisms across different timescales produces complex cognitive processes. To achieve comparable cognitive complexity in memristive neuromorphic systems, devices that are capable to emulate short- and long-term plasticity (STP and LTP, respectively) concomitantly are essential. However, this fundamental bionic trait has not been reported in any existing memristors where STP and LTP can only be induced selectively because of the inability to be decoupled using different loci and mechanisms. In this work, we report the first demonstration of truly concomitant STP and LTP in a three-terminal memristor that uses independent physical phenomena to represent each form of plasticity. The emerging layered material $Bi_2O_2Se$ is used in memristor for the first time, opening up the prospects for ultra-thin, high-speed and low-power neuromorphic devices. The concerted action of STP and LTP in our memristor allows full-range modulation of the transient synaptic efficacy, from depression to facilitation, by stimulus frequency or intensity, providing a versatile device platform for neuromorphic function implementation. A recurrent neural circuitry model is developed to simulate the intricate "sleep-wake cycle autoregulation" process, in which the concomitance of STP and LTP is posited as a key factor in enabling this neural homeostasis. This work sheds new light on the highly sophisticated computational capabilities of memristors and their prospects for realization of advanced neuromorphic functions.

Synaptic plasticity is the cellular foundation of memory and learning.[1] There are two types of plasticity that are concomitant but act over different timescales: short-term plasticity (STP), which modulates the synaptic transmission efficacy dynamically at the conveying transients but leaves the efficacy unchanged during steady-state transmission; and long-term plasticity (LTP), which, in contrast, renders stable changes in the synaptic transmission efficacy. The STP and LTP have different computational uses: STP has profound effects on motor control, speech recognition and working memory, while LTP is essential to encoding of spatial information.[2] In many if not all cases, a single plasticity is not sufficient to account for the intricate developmental and learning mechanisms, and orchestration of the STP and LTP is required to support the maximally adaptive behaviour and sophisticated cognitive functions.[3,4]

Inspired by the neural mechanisms, the semiconductor research community has envisioned new computational capabilities based on neuromorphic computing with the aim of charting a new path beyond the decades-old approach to computing based on the Von Neumann architecture as implemented with transistor-based processors.[5,6] Since the seminal discovery of the memristive behavior, which had been predicted for use in emulation of synaptic plasticity, extensive studies of memristors have been performed using various material systems, where strengthening and weakening of the synaptic connection have been mimicked using the electrically, magnetically, thermally or optically tunable resistance of the materials.[7-9] STP and LTP have been demonstrated selectively in different devices: STP has been achieved using the device volatility, where a temporal change in the resistance decays rapidly to the baseline value, while LTP has been achieved via a more prolonged change in the resistance. Irreversible transitions from STP to LTP have also been demonstrated, thus emulating the memory consolidation.[10-17] However, the basic trait of biological synapses where the STP and the LTP are concomitant has not yet been demonstrated in any existing devices.

Recently, crossbar arrays of memristors that provide network complexity have been used to perform basic cognitive functions, including pattern recognition, which were solely reliant on the steady changes in the device resistance (LTP).[18-22] Use of the computational potential of STP is expected to improve the capability of these arrays to perform higher-level cognitive functions. However, this remains challenging because the STP and LTP are induced selectively rather than concomitantly in existing devices. In biological synapses, it is widely accepted that the STP and LTP are expressed using different subcellular mechanisms over different timescales. For example, in CA1 synapses in the mammalian hippocampus, LTP is dependent primarily on modification of postsynaptic NMDA receptors, while STP is caused by an increase in the probability of transmitter release at presynaptic terminal, usually due to the residual and transient elevation in presynaptic $Ca^{2+}$. [23,24] In current memristors, however, the STP and LTP are selectively induced by controlling the degree of device volatility that is virtually of the same physical origin, in contrast to biological systems. For example, Ohno et al reported the selective induction of STP and LTP in a single device, in which both types of plasticity occurred in the $Ag_2S$ electrolyte as a result of the electrochemical reaction of the $Ag^+$/Ag pair.[16] Du et al used a single state variable to mimic both the STP and LTP mechanisms.[25] In these cases, STP simply becomes a reversible form of LTP. Wang et al mimicked the STP and LTP mechanisms using the diffusion and drift of the nanoparticles, respectively, but used two separate devices.[12]

In this work, we report the first demonstration of truly concomitant STP and LTP in a three-terminal memristor. The design philosophy is as follows: during device operation, voltage bias from the metal gate terminal either attracts or repels charge carriers in the semiconducting channel, thus modulating the channel resistance. The resulting resistance state is volatile because it cannot be maintained after the gate bias ceases. This is analogous to the transient nature of STP. The gate bias can also affect the electronic structures of the gate dielectric and/or the dielectric-semiconductor interface. Phenomena including charge trapping by the border traps or ambient molecules and polarization flipping (of ferroelectric gate dielectrics) under the influence of the gate bias have been reported in numerous studies.[26-33] These changes in the electronic structures can often be sustained in the absence of a gate voltage and can thus modulate the channel resistance in a nonvolatile manner. This is analogous to the prolonged nature of LTP. Several works have reported LTP based on these mechanisms.[33,35] From these perspectives, three-terminal memristors are desirable for decoupling of the STP and LTP through use of different functional regions and physical mechanisms, thus being a naturally suitable medium in which STP and LTP can be induced concomitantly and expressed independently. In the study of three-terminal organic synaptic devices by Burgt et al., similar physical origins of the selective STP and LTP effects have been recognized.[36] Three-terminal structure has generic utility in conventional electronics. From other performance and functionality viewpoints, decoupling of the control terminal and the transduction terminals in three-terminal memristors produces improved state retention and energy efficiency and enables performance of concurrent read and write operations.[36,37] In designing the three-terminal memristor, we use the air-stable, high-mobility semiconducting layered $Bi_2O_2Se$ as the channel material, thus potentially enabling fast synaptic information processing and ultrathin device fabrication.[38] Hall mobility of $Bi_2O_2Se$ is ultrahigh ~18000 $cm^2V^{-1}s^{-1}$ at cryogenic temperature.[38] This value is comparable to what is observed in graphene grown by chemical vapour deposition and at the $LaAlO_3$–$SrTiO_3$ interface.[39,40] The room temperature Hall mobility is ~200 $cm^2V^{-1}s^{-1}$.[38] In addition to mobility advantage, its air-stability makes it a strong contender to typical Van der Waals stacked layered materials such as $MoS_2$, black phosphorus and InSe.[41-43] For the first time, $Bi_2O_2Se$ is demonstrated to be a promising material for neuromorphic computing.

A chemical synapse is illustrated schematically in **figure 1(a)**. This synapse passes information unidirectionally from a presynaptic neuron to a postsynaptic neuron and is thus asymmetric in both structure and function. The presynaptic terminal contains neurotransmitters, while the postsynaptic terminal contains neurotransmitter receptors. The arrival of a nerve impulse from the presynaptic neuron triggers the release of neurotransmitters into the cleft between the pre- and postsynaptic terminals. Some of these neurotransmitters then bind to the receptors and liaise with the postsynaptic neuron to allow the signals to be conveyed in the neural circuits. The synaptic transmission efficacy can be altered by the preceding activities, and this is known as the synaptic plasticity. As mentioned earlier, the STP and LTP occur at the pre- and postsynaptic terminals, respectively.

The cross-sectional structure of the $Bi_2O_2Se$ three-terminal memristor is shown in **figure 1(b)**. To mimic signal transmission through a synapse, the vertical gate stack of the $Bi_2O_2Se$ device acts as the presynaptic terminal that receives the presynaptic input signal in the form of the gate voltage ($V_g$), while the in-plane channel under a constant source-drain bias ($V_{ds}$=100 mV) acts as the postsynaptic terminal that transmits the postsynaptic output signal, i.e., the postsynaptic current (PSC), in the form of the source-drain current ($I_{ds}$) and in response to $V_g$. An optical image of the device is shown in **figure 1(c)**. The channel length of the device is ~500 nm. The $Bi_2O_2Se$ and $HfO_2$ layer thicknesses are 5 nm and 20 nm, respectively. The source, drain and gate electrode layers are composed of 5-nm-thick Pd and 40-nm-thick Au films (see Methods).

Because the synaptic plasticity is a history-dependent change in the synaptic strength, cyclic gate voltage sweep measurements (see **Experimental Section**) are performed to determine the effects of preceding activities on the output $I_{ds}$. **Figure 1(d)** shows the transfer characteristics of the $Bi_2O_2Se$ device measured under various peak-to-peak $V_g$ sweeps. The voltage is swept at a low rate to allow the steady-state $I_{ds}$ values to be measured. In addition to n-type behaviour, a hysteresis loop is observed, indicating that $I_{ds}$ is dependent on the history of $V_g$ sweeping history. Forward sweeps (negative to positive) and backward sweeps (positive to negative) lead to positive and negative shifts in the transfer curves, respectively. This indicates that a positive (negative) $V_g$ leads to suppression (enhancement) of the steady-state channel conductivity despite its transient enhancement (suppression). As expected, the magnitude of this type of shift is determined by the intensity of the applied $V_g$. The long-lasting modulation capability of the steady-state channel conductivity can thus be used to emulate the LTP of synapses. The basic device properties, including operation speed, endurance and retention, are presented in **figure S1**.

Based on screened exchange (sX-LDA) hybrid density functional calculations (see **Experimental Section**), we propose a charge trapping/detrapping mechanism that could account for the hysteresis transfer characteristics of our device. sX-LDA functional produces band gap values of ~0.6 eV and ~6.0 eV for bulk $Bi_2O_2Se$ and $HfO_2$, respectively, in close agreement with experimental values. **Figure 1(e)** shows the atomic structure model of a $Bi_2O_2Se$ (Bi-terminated): $HfO_2$ (O-terminated) interface. The projected atomic partial density of states (PDOSs) of the interface are shown in **figure 1(f)**. The interface is found to be insulating, thus satisfying the device requirement. Valence and conduction band offset values of 4.0 eV and 1.4 eV, respectively, are deduced from the PDOSs. Because the $HfO_2$ gate dielectric is deposited at a relatively low temperature, the concentration of the randomly distributed fixed charges is expected to be high in $HfO_2$. Oxygen vacancy (Ov) is commonly considered as a fixed charge center in $HfO_2$. Based on the previously reported defect charge-state transition levels, the most stable defect states are in the sequence $Ov^{+1}$ and $Ov^0$ for increasing Fermi energy when the Fermi energy is taken to be in the $Bi_2O_2Se$ band gap and the calculated band offsets are used.[44] Specifically, the transition level $E_{+1/0}$ is situated close to the conduction band minimum of $Bi_2O_2Se$. For $Ov^{+1}$, occurrence of unoccupied electron states in $HfO_2$ band gap enables electron trapping (detrapping) in (from) $HfO_2$ by which the addition (loss) of an electron causes a transition from $Ov^{+1}$ ($Ov^0$) to $Ov^0$ ($Ov^{+1}$).[45-47] This provides an electron trapping/detrapping energy level corresponding to $E_{+1/0}$. The proposed charge trapping/detrapping mechanism is shown schematically in **figure 1(g)**. It is worth pointing out that the charge trapping/detrapping mechanism is inherently slow compared with other mechanisms, such as ferroelectric switching which could be used to boost the operation speed of our device toward nanosecond scale comparable to state-of-the-art high speed two-terminal memristors.[17, 48-52]

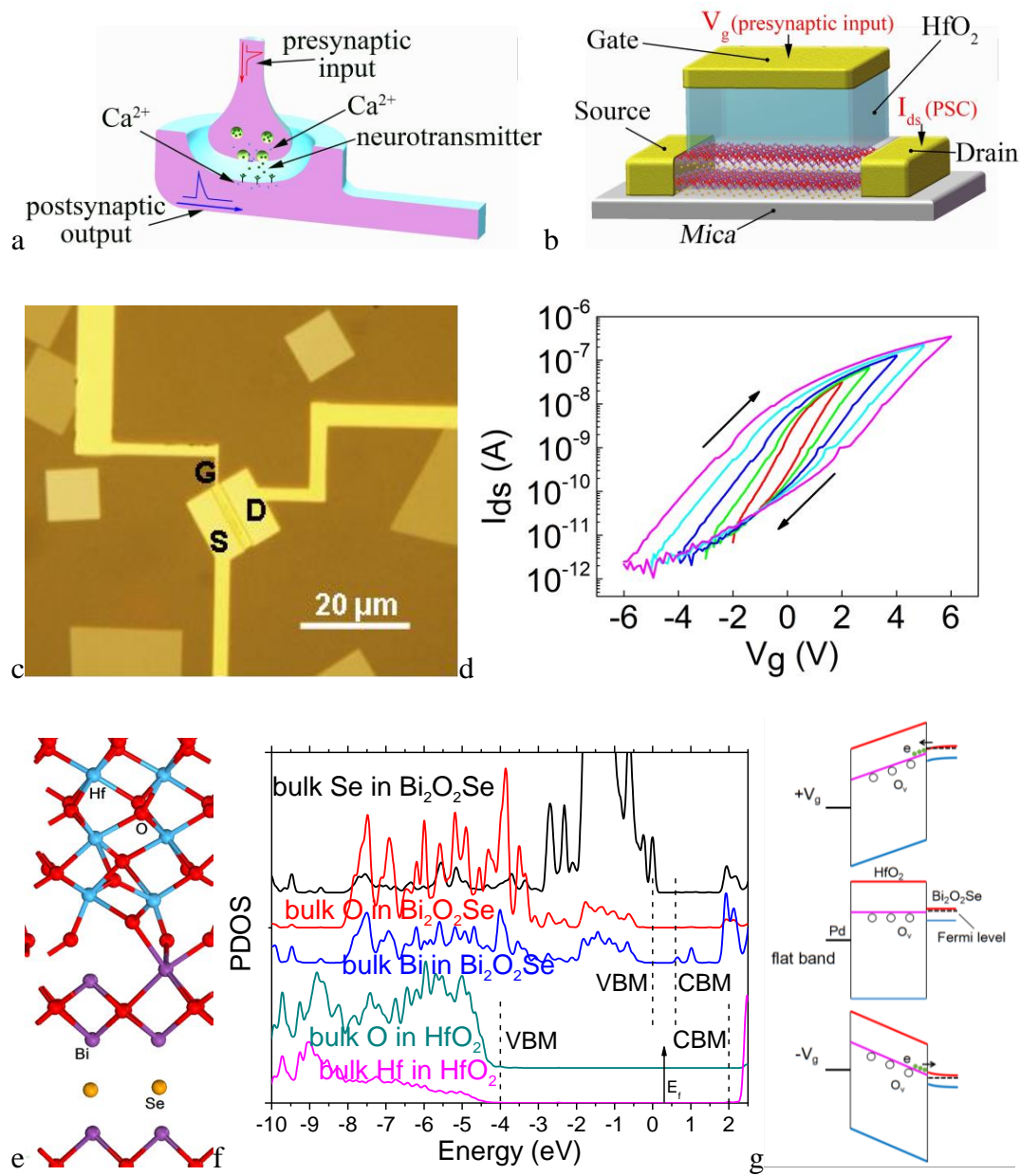

**Figure 1 (a)** Schematic of a chemical synapse. **(b)** Cross-sectional structure of the Bi$_2$O$_2$Se three-terminal memristor. **(c)** Optical image of the Bi$_2$O$_2$Se three-terminal memristor. **(d)** Hysteretic transfer curves for various peak-to-peak V$_g$ sweeps. **(e)** Atomic structure of Bi$_2$O$_2$Se: HfO$_2$ interface**. (f)**  Projected atomic PDOSs of the interface. **(g)** Schematic of the charge trapping/detrapping mechanism. The Fermi energy of Pd is aligned to the band edges of HfO$_2$ using the experimental work function of Pd and electron affinity of HfO$_2$, while the band edges of Bi$_2$O$_2$Se is aligned to the band edges of HfO$_2$ using the calculated values. Ov induced trapping level E(+1/0) is positioned according to the results in ref. 44.

To simulate a synapse in a spiking neural network, we first perform single pulse measurements (see **Experimental Section**). **Figure 2(a)** shows the PSC (=$I_{ds}$) of the $Bi_2O_2Se$ memristor when stimulated using a single triangular positive voltage pulse (+5 V, 50 ms) as the presynaptic input signal. The constant bias $V_{ds}$ across the channel is 100 mV and the base $I_{ds}$ value is ~3 nA before arrival of the voltage pulse. Application of a positive voltage pulse results in a PSC as high as 133 nA because of transient accumulation of electrons in the channel. After the pulse elapses, the PSC decreases to a lower base value of ~1 nA. This indicates that the channel becomes less conductive, i.e., it becomes less n-type or more electron are depleted. This is consistent with the quasi-DC voltage sweep measurement results. This effect from application of the positive $V_g$ is analogous to synaptic long-term depression ($LT_d$), which is a special type of LTP. In contrast, long-term potentiation ($LT_p$), which is another type of LTP, can be mimicked using a negative voltage pulse (see **figure S2(a)**), which is consistent with the quasi-DC voltage sweep measurement results. The dependence of the peak PSC value ($PSC_{max}$) on the intensity of the positive gate voltage pulse is shown in **figure 2(b)**. The peak value increases almost linearly from ~ 3 nA to ~282 nA with increasing pulse intensity from 1 V to 10 V as a result of enhanced electron accumulation in the channel. The dependence of the base $I_{ds}$ value ($I_{ds}^{base}$) after $V_g$ elapses on the pulse intensity is also illustrated. In contrast to $PSC_{max}$, $I_{ds}^{base}$ decreases with increasing $V_g$. The dependence of $I_{ds}^{base}$ on the intensity of the negative gate voltage pulse are shown in **figure S2(b)**. Compared with $LT_d$ effect, $LT_p$ under negative gate voltage is weaker. We attribute this asymmetry to electron accumulation under positive gate voltage that enhances transient electron charging into the $HfO_2$ layer, as in the case of bias temperature instability for typical high-K oxide based metal-oxide-semiconductor field-effect transistors (MOSFETs).[48, 53, 54] This could be even more significant in our $Bi_2O_2Se$ device because of its strong n-type properties.[38]

Unlike the LTP, the STP acts primarily to modify the synaptic transmission efficacy dynamically at the conveying transients but leaves the efficacy unchanged during steady-state transmission. We perform pulse train measurements (see Methods) and find that the transient synaptic efficacy can be modulated over the full-range from depression to facilitation via the concerted action of the STP and LTP. **Figure 2(c)** shows the PSC that is triggered using a train of $V_g$ pulses (+1 V, 50 ms). The interval between two pulses is 50 ms. The results show that $PSC_{max}$ decreases with increasing numbers of pulses. This can be expected as a result of the $LT_d$ effect of each positive $V_g$ pulse. When the pulse intensity increases to +4 V, the $PSC_{max}$ that is triggered by each pulse becomes almost identical, as shown in **figure 2(d)**. This is quite unexpected, simply because stronger pulses should cause a more pronounced $LT_d$. This observation can only be rationalized by taking the short-term facilitation ($ST_f$), a special type of STP, into consideration. As mentioned earlier, the accumulated electrons in the channel that are induced by a positive $V_g$ pulse are recovered over a short period after the pulse elapses. If a subsequent pulse arrives before the electrons recover completely, residual electrons may compensate the depressed electron accumulation because of the $LT_d$ effect of the preceding pulse; in other words, the $LT_d$ is balanced by the $ST_f$ at the conveying transient. Accordingly, for a +1 V pulse, the number of accumulated electrons induced by each pulse is small so that they can be recovered completely on the arrival of the subsequent pulse. Therefore, the $LT_d$ effect is predominant, which results in the depressed synaptic efficacy. This hypothesized mechanism is supported by the observation that $PSC_{max}$ increases with increasing numbers of pulses during the +10 V pulse train measurements, as shown in **figure 2(e)**; this behaviour can be explained as the overcompensation of the $LT_d$ by the $ST_f$. We define $\Delta_n = \frac{(PSC_{max}^n - PSC_{max}^1)}{PSC_{max}^1}$ as the percentage change in the n*th* $PSC_{max}$ relative to the first $PSC_{max}$. **Figure 2(f)** shows $\Delta_n$ as a function of the pulse number and its dependence on the pulse intensity. The figure shows that when the pulses are weak, $PSC_{max}$ decreases with increasing numbers of pulses. Increasing the pulse intensity leads to a lower rate of decrement or even an increment in PSC$_{max}$ as the number of pulses increases. The pulse intensity-dependent working mechanisms of the device under the specific 50 ms pulse interval condition are shown pictorially in **figure S3**. Significant short-term effect at a timescale longer than the normal dielectric relaxation time may indicate the involvement of certain slow ion processes that mediate the channel modulation. In this regard, the movement of interfacial Se anions are most likely because they are stabilized by weak electrostatic interaction and Se vacancies are shallow n-type donors.[55] The results of the alternating positive/negative V$_g$ pulse train measurements are shown in **figure S4**, showing symmetric and stable multi-level resistive switching behaviour. Lower-voltage operation of our device is shown in **figure S5**. ST$_f$/LT$_d$ and ST$_d$/LT$_p$ effects are observed under positive and negative pulse train measurements with low pulse intensity of 10 mV and 100 mV, respectively, comparable to state-of-the-art low-voltage synaptic devices. [36] This implies the potential for low-power neuromorphic applications. Compared with ST$_f$ effect, ST$_d$ under negative gate voltage is less significant (higher pulse intensity required). We attribute this asymmetry to more (less) accumulated (depleted) electrons under the preceding positive (negative) gate bias of a given intensity and therefore less (more) complete recovery of channel carrier density on the arrival of the subsequent pulse in a given time interval, as can be deduced from the transconductance of our n-type device in the depletion mode (see **figure 1(d)**).

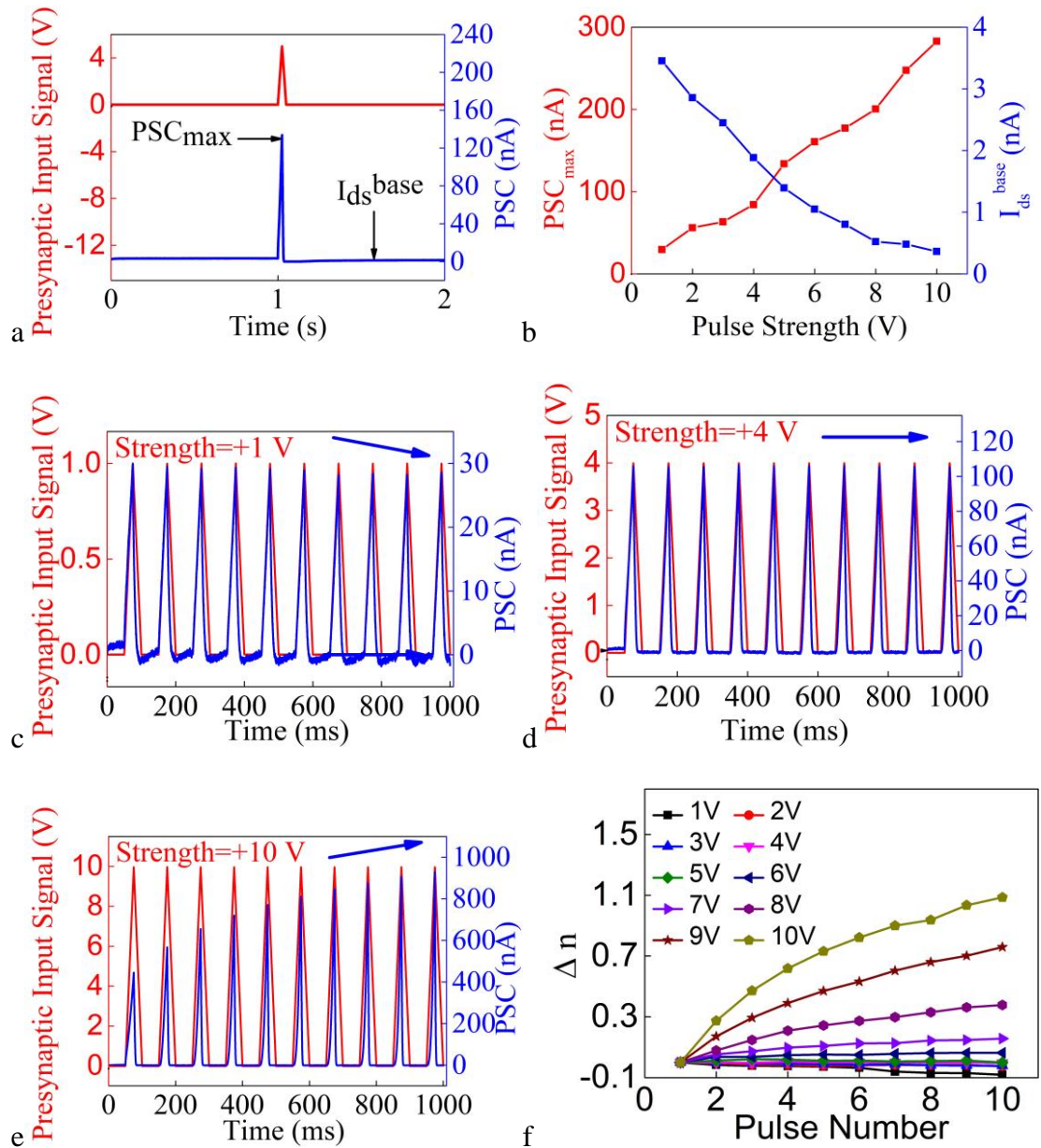

**Figure 2 (a)** PSC of the Bi$_2$O$_2$Se synaptic device when triggered using a presynaptic input signal (+5 V, 50 ms). **(b)** Dependences of PSC$_{max}$ and I$_{ds}^{base}$ on the intensity of the stimulating pulse. **(c)-(e)** PSC triggered using a train of V$_g$ pulses with a pulse interval of 50 ms, where the pulse intensities are +1 V, +4 V and +10 V for (c), (d) and (e), respectively. **(f)** $\Delta_n$ as a function of pulse number and its dependence on the pulse intensity. The pulse width is fixed at 50 ms.

In neuroscience, study of the frequency-dependent changes in synaptic efficacy is essential to unveiling of the mechanisms that underlie neuronal information processing. Similarly, measurements are performed here using pulse trains at various frequencies to investigate the symphonic effects of the STP and LTP in our device. We first apply a train of $V_g$ pulses with low intensity of +2 V. The results show that $PSC_{max}$ always decreases with increasing numbers of pulses, regardless of the frequency of the pulse stimuli, as shown in **figure 3(a1)-(a3)**. This can be understood based on the predominance of the $LT_d$ effect over the $ST_f$ effect: pulses with low intensity induce small numbers of accumulated electrons that can be recovered completely on the arrival of the subsequent pulse, even when the interval is short (high frequency); without the aid of the $ST_f$, the synaptic efficacy can only be depressed. The percentage change in the n*th* $PSC_{max}$ relative to the first $PSC_{max}$ is shown in **figure 3(a4)** for each pulse interval.

An increase in the intensity of the pulses causes different frequency-dependent synaptic transmission behaviour. When we apply a train of $V_g$ pulses with intensity of +6 V, we observe a transition from depression to facilitation of the transient synaptic efficacy when the frequency of the stimuli increases, as shown in **figure 3(b1)-(b3)**. As expected, $PSC_{max}$ decreases with increasing numbers of pulses at low frequencies (interval=1 s) because of the predominant $LT_d$ effect. When the frequency increases (interval=200 ms), we observe that $PSC_{max}$ remains almost unchanged because of the compensation of the $LT_d$ by the $ST_f$. At even higher frequencies (interval=10 μs), the $ST_f$ becomes dominant over the $LT_d$ and therefore an increase in $PSC_{max}$ with increasing numbers of pulses is observed. **Figure 3(b4)** shows the percentage change in the n*th* $PSC_{max}$ relative to the first $PSC_{max}$ for each pulse interval. A transition from depression to facilitation is clearly observed as the frequency increases. The timescale of this $ST_f$ effect is estimated to be hundreds of milliseconds (see **figure S6**). The pulse interval-dependent working mechanisms of the device under this specific +6 V pulse intensity condition are shown pictorially in **figure S7**.

A further increase in pulse intensity to +10 V leads to new frequency-dependent synaptic transmission behaviour. From **figure 3(c1)-(c3)**, we see that PSC$_{max}$ increases with increasing numbers of pulses, even under low frequency conditions. This is attributed to the strongly enhanced ST$_f$. The strong V$_g$ pulse produces large numbers of residual electrons that are instantaneously available on the arrival of the subsequent pulse. **Figure 3(c4)** shows the percentage change in the PSC$_{max}$ as a function of the number of pulses. The figure shows that the transient synaptic efficacy is always facilitated over the range of the frequencies tested.

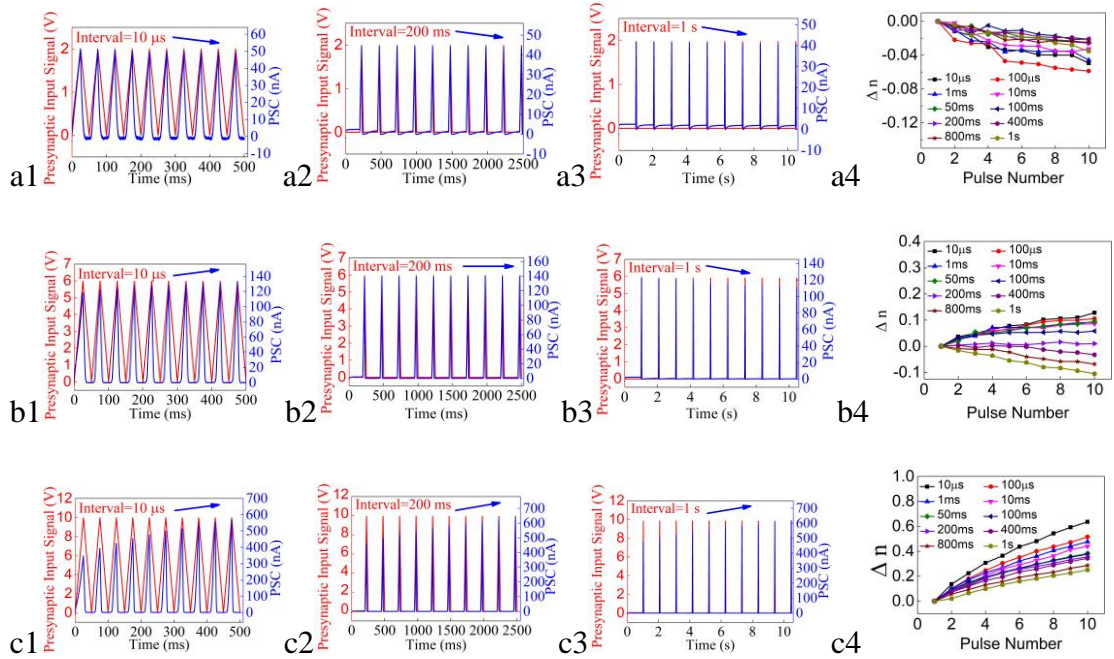

**Figure 3 (a1-a3)** PSC triggered using a train of $V_g$ pulses with the pulse intensity of +1 V, where the pulse intervals are 10 $\mu s$, 200 ms and 1s for (a1), (a2) and (a3), respectively. **(a4)** $\Delta_n$ as a function of pulse number and its dependence on the pulse interval. **(b1-b3)** PSC triggered using a train of $V_g$ pulses with the pulse intensity of +6 V, where the pulse intervals are 10 $\mu s$, 200 ms and 1s for (b1), (b2) and (b3), respectively. **(b4)** $\Delta_n$ as a function of pulse number and its dependence on the pulse interval. **(c1-c3)** PSC triggered by a train of $V_g$ pulses with pulse intensity of +10 V, where the pulse intervals are 10 $\mu s$, 200 ms and 1s for (c1), (c2) and (c3), respectively. **(c4)** $\Delta_n$ as a function of pulse number and its dependence on the pulse interval. The pulse width is fixed at 50 ms.

To demonstrate the sophisticated computational capabilities of our memristor, which uses its orchestrated LTP and STP to enable full-range modulation of the transient synaptic efficacy from depression to facilitation by stimulus frequency, the intricate neural process that underlies "sleep-wake cycle autoregulation" is simulated. Despite of the importance of this neural function, the mechanisms behind it are only partially understood. Inspired by a recent finding in fruit flies that recurrent neural circuitry exists to balance sleep need, we propose a heuristic recurrent neural circuit model (**figure 4a**) in which the concomitance of the synaptic LTP and STP is posited as a key factor in enabling this neural homeostasis.[56] The model consists of two neurons (Ne1, Ne2) and two synapses (Sy1, Sy2). Ne1 may receive external stimuli via Sy1, such as sunlight and sensory signals indicating enemy invasion, to trigger locomotion, or may receive internal sleeping signals from Ne2 to inhibit locomotion. In turn, Ne2 may receive signals from Ne1 via Sy2 to generate sleeping signals.

A potential electronic circuit implementation of the above neural circuit is shown in **figure 4b**. In addition to the $Bi_2O_2Se$ synaptic devices, other circuit elements required for a fully-fledged operation include resistors, capacitors, diodes, MOSFET switches and comparators. The circuit includes two electronic synapses (Sy1 and Sy2) and two electronic neurons (Ne1 and Ne2). External stimuli for regulation of sleep, such as sunlight and the approach of natural enemies, are transduced into positive presynaptic voltage inputs. Sy1 and Sy2 show $LT_d$ and $ST_f$ concomitantly under the application of a positive gate voltage, and it is expected that they will conversely show $LT_P$ and $ST_d$ under a negative gate voltage. Ne1 is specifically designed to have dual firing thresholds with opposite polarities so that it fires if the membrane potential exceeds either of the two thresholds. In addition, the polarity of the Ne1 firing spikes is reversed before they reach Sy2 so that the expressed plasticity of Sy2 is always opposite to that of Sy1. Similar to Ne1, Ne2 also has dual thresholds with opposite polarities. Unlike Ne1, however, the Ne2 firing spikes are unipolar so that they always counteract the external stimuli.

The "sleep-wake cycle autoregulation" process is simulated (**figure 4c**) using a recurrent spiking neural network. Specifically, the external sunlight stimulus changes chronologically from strong during the daytime, to medium as dusk approaches, and eventually becomes low at midnight; these changes are represented here by changes in frequency. During the daytime, a strong and positive sunlight stimulus causes lasting integration-and-firing of Ne1 at high frequency because of the strong initial Sy1 synaptic connection and the $ST_f$ effect on Sy1, despite the concomitant $LT_d$ effect. The importance of the $ST_f$ is specifically manifested in the case of enemy invasion, particularly at night, when immediate wake-up and locomotion is essential (see **figure S8**). The $LT_d$ effect is analogous to fatigue of the organism. Firing of Ne1 maintains the daytime locomotion. However, because Sy2 receives high-frequency negative Ne1 firing spikes, the $ST_d$ effect prevents the spikes from passing and cuts off the path through Sy2. Simultaneously, the concomitant $LT_p$ effect on Sy2 causes strengthening of the steady-state synaptic connection.

At dusk, however, the moderate-intensity sunlight stimulus (medium frequency) cannot induce the $ST_f$ effect on Sy1. Additionally, the steady-state Sy1 synaptic connection is very weak because of $LT_d$ during the daytime. Therefore, Ne1 only fires at lower frequencies. This allows Ne1 firing spikes to pass through Sy2 because of the insignificant $ST_d$ effect. Given the strong steady-state Sy2 synaptic connection caused by $LT_p$ during the daytime, Ne2 can then integrate the Sy2 postsynaptic output and fire, thus establishing a recurrent circuit. Because the sunlight stimulus becomes much weaker at dusk, the counteracting Ne2 firing spikes reverse the expressed plasticity of Sy1 to $LT_p$ and $ST_d$ (weak), leading to gradual recovery of the steady-state Sy1 synaptic connection. Consequently, the polarity of the Ne1 firing spikes is also reversed, thus inhibiting locomotion; in other words, sleep is induced. Simultaneously, the expressed plasticity of Sy2 is also reversed to $LT_d$ and $ST_f$ (weak), leading to gradual recovery of the steady-state Sy2 synaptic connection. This is analogous to functional restoration or self-healing of organisms during sleep.

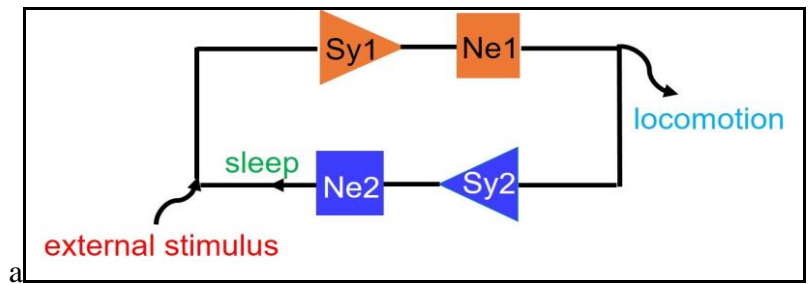

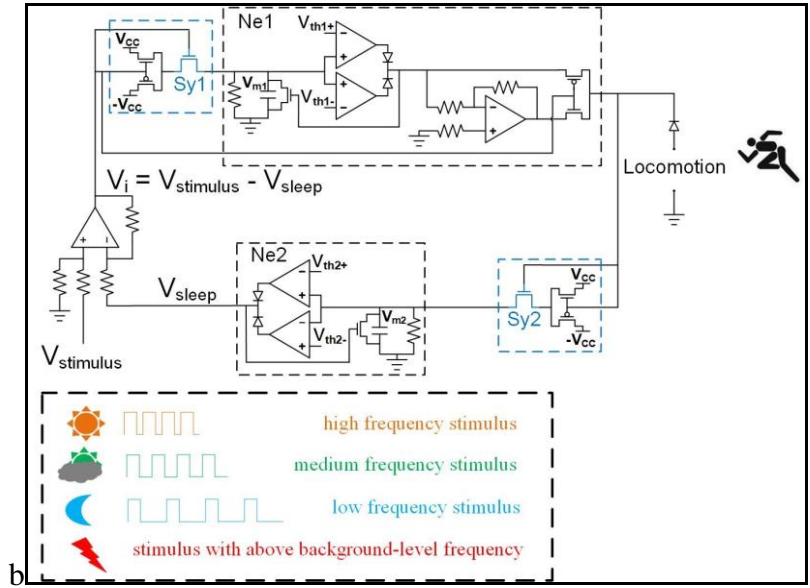

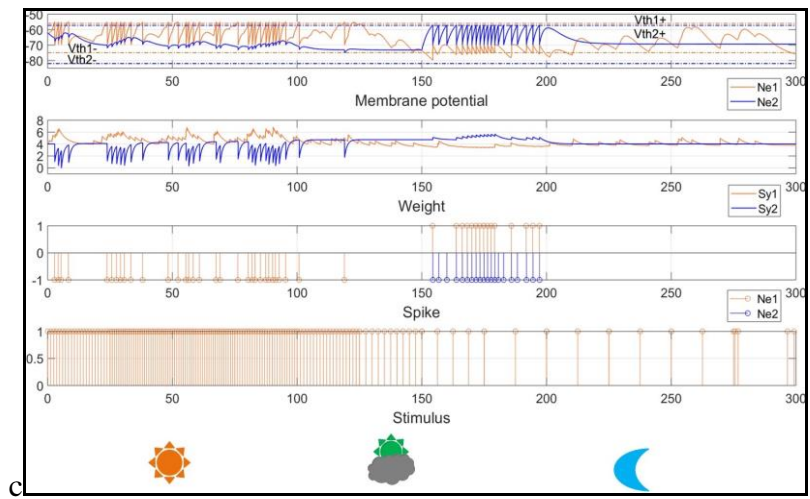

**Figure 4 (a)** Heuristic recurrent neural circuit model for "sleep-wake cycle autoregulation". **(b)** Electronic implementation of the recurrent neural circuit. **(c)** Recurrent spiking neural network simulation of "sleep-wake cycle autoregulation". The changes in membrane potentials of Ne1 and Ne2, in synaptic weights of Sy1 and Sy2, and the spiking activities of Ne1 and Ne2 in response to external stimuli are presented.

In conclusion, concomitant LTP and STP are demonstrated for the first time in our three-terminal $Bi_2O_2Se$-based memristor; this is achieved by decoupling the loci at which the physical LTP and STP processes occur and by capitalizing on their different physical mechanisms. This is fundamentally different from existing memristor behaviour and represents a significant device improvement for capture of the inherent biological attributes of synapses. The concerted action of the STP and LTP in our memristor allows full-range modulation of the transient synaptic efficacy from depression to facilitation by stimulus frequency or intensity, thus providing a versatile device platform for neuromorphic function implementation. Use of the emerging layered material $Bi_2O_2Se$ in the memristor offers the potential for realization of ultrathin, high-speed and low-power devices. To demonstrate the sophisticated computational capability of this memristor with its orchestrated LTP and STP, the intricate neural process that underlies "sleep-wake cycle autoregulation" is simulated using a heuristic recurrent neural circuitry model, in which the concomitance of STP and LTP is posited as the key factor in enabling this neural homeostasis. By providing the memristor family with an important addition that has been required for a decade, this work sheds new light on the highly sophisticated computational capabilities of memristors and their prospects for realization of advanced neuromorphic functions.

**Experimental Section**

*Device fabrication:* $Bi_2O_2Se$ memristors are fabricated on chemical vapour deposited (CVD) samples with anaverage thickness of ~5nm. The source and drain contacts are patterned into the poly (methyl methacrylate) (PMMA, A4, 200 nm thick) using electron beam lithography (EBL). 5-nm-thick Pd and 40-nm-thick Au layers are deposited by thermal evaporation followed by a lift-off process in 80°C acetone. A second EBL step is then performed to pattern the gate electrode. A 20-nm-thick $HfO_2$ layer to serve as the gate dielectric is deposited by thermal atomic layer deposition (TALD) at 70°C. TALD is performed at a relatively low temperature, which increases the concentration of the randomly distributed fixed charges in the gate oxide. Pd/Au electrodes are then formed by lift-off of the thermally evaporated 5-nm-thick Pd and 40-nm-thick Au layers to complete the fabrication process. All devices have the same channel length of ~500 nm.

*Electrical characterization:* Electrical characterization is performed using a Keysight B1500A semiconductor cell analyzer, which is supplemented with a B1530A waveform generator/fast measurement unit (WGFMU) that is used to perform the pulse measurements. All electrical measurements are performed at room temperature and under an ambient atmosphere.

*Hybrid density functional calculation:* Calculations are performed using the Cambridge Sequential Total Energy Package. 86 atoms' interface supercell model is used. Cutoff energy of the plane wave basis set is 680 eV. After geometry optimization using the generalized gradient approximation, the sX-LDA hybrid density functional is applied to the relaxed atomic structure to obtain the accurate interface band structures. The sX-LDA functional is closely related to the Hartree-Fock (HF) method. This functional is an implicit functional of the density and part of the exchange-correlation (XC) operator is nonlocal. The nonlocal XC potential is similar in form to the HF potential but it also incorporates the effects of correlation by screening the long-range interactions of exchange. This is achieved by introducing a factor which decays exponentially with electron separation. The short-range component of the exchange contribution is thus given by

$$E_x^{HF,SR} = -\sum_{i,j} \iint \frac{\psi_i^*(r)\psi_i(r') \exp^{-k_{TF}|r-r'|})\psi_j^*(r')\psi_j^*(r)}{|r-r'|} d^3r' d^3r$$

where $k_{TF}$ is the reciprocal Thomas-Fermi screening length. However, it is advantageous to maintain the exact XC energy for the homogeneous electron gas (HEG). Therefore, a local contribution is also required so that the total XC energy in the screened exchange method is

$$E^{sX} = E_{nl}^{sX} + E_{loc}^{sX}$$

where the local part is parameterized using Perdew's expression for the local density approximation (LDA). As a result, sX-LDA ensures that both the LDA and HF limits are correct for the HEG at long and short screening lengths. sX-LDA improves the calculated band gap of a wide range of semiconductors and insulators compared to the LDA.

*Spiking neural network simulation:* To establish the neural dynamics, we adopt the leaky integrate-and-fire (LIF) model. The membrane voltages $V_1$ and $V_2$ for Ne1 and Ne2, respectively, are given by

$$\tau \frac{dV_1}{dt} = (E_{rest} - V_1) + k_g g_1 (E_g - V_1) + I_1 \sum_{t_i} \delta(t - t_i)$$
$$\tau \frac{dV_2}{dt} = (E_{rest} - V_2) + k_g g_2 (E_g - V_2) + I_2 \sum_{t_i} \delta(t - t_i)$$

where $E_{rest}$ is the resting membrane potential that is common to both neurons, $I_1$ and $I_2$ are the corresponding input currents, $\tau$ is the membrane voltage decay constant that is common to both neurons, $k_g$ is the conductance constant that is common to both neurons, $t_i$ is the time when the *i*th firing occurs, $E_g$ is the equilibrium membrane potential that is common to both neurons, and $g_1$ and $g_2$ are the conductances of Sy1 and Sy2, respectively. These conductances, $g_1$ and $g_2$, respectively, are described by

$$\tau_1 \frac{dg_1}{dt} = -g_1 + k_1^s \sum_{t_i} \delta(t - t_i) + k_1^L \sum_{t_i} H(t - t_i)$$
$$\tau_2 \frac{dg_2}{dt} = -g_2 + k_2^s \sum_{t_i} \delta(t - t_i) + k_2^L \sum_{t_i} H(t - t_i)$$

where $\tau_1$ and $\tau_2$ are the corresponding conductance constants, respectively, and $k_1^S$, $k_1^L$, $k_2^S$ and $k_2^L$ are the corresponding strengths of the synapse plasticity, in which S and L denote the STP and LTP, respectively. The Heaviside function $H(x)$ and the Dirac function $\delta(x)$ are used to describe the effects of the LTP and STP, respectively Upon reception of a firing spike, the synaptic conductance will be either transiently facilitated ($ST_f$) or depressed ($ST_d$). This short-term effect is manifested in the Dirac function $\delta(x)$. Additionally, because of the concomitant LTP, the synaptic conductance will also be steadily potentiated ($LT_p$) or depressed ($LT_d$) to a certain extent. This long-term effect is handled using the Heaviside function. Because of the existence of the short-term effects, the synaptic conductance, after modulation, will gradually decay to a new baseline value that is determined by the LTP effect. Unlike biological neurons and therefore unlike the canonical LIF neuron models, our neuron model has dual thresholds and a reversed spike polarity (for Ne1), as described in the main text. A membrane potential that exceeds either of the two thresholds, denoted by $th_+$ and $th_-$, will cause firing of the neuron. After firing, the neuron membrane potential is reset to $E_{rest}$. We used the Euler method to solve the required differential equations. The values of the different parameters used in the simulations are listed in **table 1**.

**Table 1** Values of the parameters used in our simulations.

| Parameter | Value | Parameter | Value |
|---|---|---|---|
| *Simulation time* | *500* | *dt* | *0.005* |
| $E_{rest}$ | *-70* | $\tau$ | *4.8* |
| $E_g$ | *0* | $k_g$ | *0.5* |
| $V_{1+}^{thr}$ | *-55* | $V_{1-}^{thr}$ | *-76* |
| $V_{2+}^{thr}$ | *-57* | $V_{2-}^{thr}$ | *-82* |
| $I_{1+}$ | *0.4* | $I_{1-}$ | *-0.4* |
| $I_{2+}$ | *2* | $I_{2-}$ | *-2* |
| $\tau_{1+}$ | *3* | $\tau_{1-}$ | *1* |
| $k_{1+}^L$ | *-0.005* | $k_{1-}^L$ | *0.005* |
| $k_{1+}^S$ | *0.6* | $k_{1-}^S$ | *-0.01* |
| $\tau_{2+}$ | *3* | $\tau_{2-}$ | *1* |
| $k_{2+}^L$ | *-0.01* | $k_{2-}^L$ | *0.025* |
| $k_{2+}^S$ | *3* | $k_{2-}^S$ | *-3* |

**Acknowledgements**

H. Li thanks National Natural Science Foundation (No. 61704096) and Beijing Natural Science Foundation (No. 4164087) for financial support. L. P. Shi thanks National Natural Science foundation (No. 61603209, 61475080, 61327902), SuZhou-Tsinghua innovation leading program (2016SZ0102) and Beijing Innovation Centre for Future Chip for financial support. H. Li thanks Guanghan Wang for drawing figure 1(a,b).

**Conflict of Interest**

The authors declare no conflict of interest.

**Author contributions**

H. Li conceived this project and designed experiments. H. Li and L. P. Shi supervised the electrical measurements. H. Li, L. P. Shi and W. Zhang supervised the spiking neural network simulations. H. Peng supervised the device fabrication. Z. Zhang, Y. He performed the electrical measurements. T. R. Li fabricated the device. Y. Wu conducted the spiking neural network simulations. Y. Jia conceived the spiking neural network simulations. C. Tan synthesized $Bi_2O_2Se$. H. Li, X. Xu and J. Lv performed the density functional calculations. G. Wang designed the electronic circuit model. H. Li wrote the manuscript. All authors contributed to the discussions and preparation of the manuscript.

*Supplementary Information*

# Truly concomitant and independently expressed short- and long-term plasticity in Bi$_2$O$_2$Se-based three-terminal memristor

*Ziyang Zhang[1+], Tianran Li[2+], Yujie Wu[1+], Yinjun Jia[3+], Congwei Tan[2], Xintong Xu[4], Guanrui Wang[1], Juan Lv[1], Wei Zhang[3*], Yuhan He[5], Luping Shi[1*], Hailin Peng[2*], Huanglong Li[1*]*

[1]Department of Precision Instrument, Center for Brain Inspired Computing Research, Tsinghua University, Beijing, 100084, China [2]Center for Nanochemistry, Beijing Science and Engineering Center for Nanocarbons, Beijing National Laboratory for Molecular Sciences, College of Chemistry and Molecular Engineering, Peking University, Beijing 100871, China [3]School of Life Sciences, Tsinghua-Peking Joint Center for Life Sciences, IDG/McGovern Institute for Brain Research, Tsinghua University, Beijing 100084, China. [4]School of Aerospace Engineering, Tsinghua University, Beijing, 100084, China [5]Department of Electronic Engineering, Tsinghua University, Beijing, 100084, China
[+]These authors contributed equally.
[*]e-mails: zhangweilab@biomed.tsinghua.edu.cn
      lpshi@mail.tsinghua.edu.cn
      hlpeng@pku.edu.cn
      li_huanglong@mail.tsinghua.edu.cn

## S1. The endurance, retention and operation speed of Bi$_2$O$_2$Se synaptic device

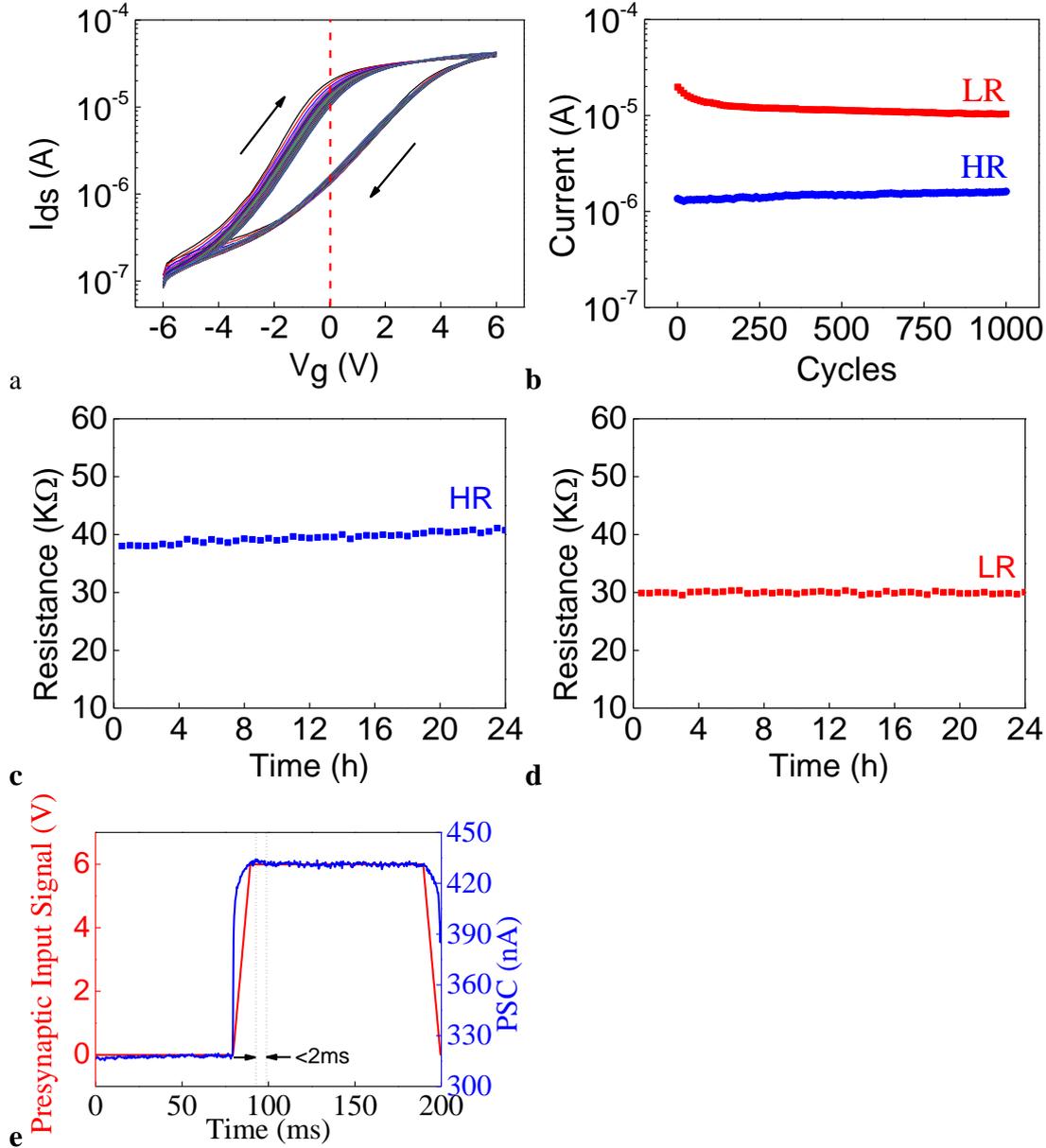

**Figure S1** (a) Hysteretic transfer curves of 1000 consecutive sweep cycles (b) The lower resistance (LR) state and higher resistance (HR) state of the Bi$_2$O$_2$Se three-terminal memristor in 1000 consecutive sweep cycles. (c) The 24-hour retention of the HR of the Bi$_2$O$_2$Se three-terminal memristor. (d) The 24-hour retention of the LR of the Bi$_2$O$_2$Se three-terminal memristor. Newly fabricated device (device 2) is used for (a-d). (e) The operation speed measurement of the Bi$_2$O$_2$Se three-terminal memristor. Newly fabricated device (device 3) is usded. The characteristic time of switching is determined by the time interval between the peak of I$_{ds}$ and the point where I$_{ds}$ decays to a steady value, analogous to the typical bias temperature instability (BTI) measurements. The measured timescale is comparable to that of fast charge injection into the high-K oxide layer in MOS stack in BTI measurements.

## S2. Mimicry of long-term potentiation (LT$_p$) using negative voltage pulse.

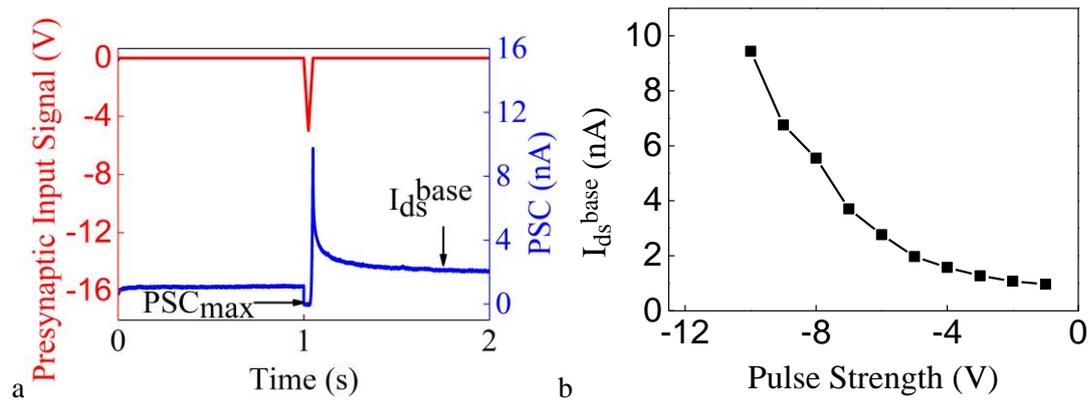

**Figure S2** (a) Postsynaptic current (PSC) of Bi$_2$O$_2$Se synaptic device when triggered using a presynaptic input signal (-5 V, 50 ms). (b) Dependence of I$_{ds}^{base}$ on the intensity of the stimulating pulse.

**S3. Pulse intensity-dependent working mechanisms of $Bi_2O_2Se$ synaptic device under the specific 50 ms pulse interval condition.**

When a positive gate voltage pulse is applied, free electron carriers are accumulated, thus increasing the channel conductivity. After the gate voltage pulse elapses, the channel conductivity gradually decreases as a result of the recovery of the steady-state carrier density. In the strong pulse intensity case, the channel conductivity on the arrival of the next gate voltage pulse may be higher than that during the preceding pulse, and this is attributed to the presnece of residual electrons. A stronger pulse causes more electrons to be accumulated and therefore more residual electrons are present. This is the electronic basis of the $ST_f$. In addition to this transient effect, the possible charge trapping in the border traps, which occurs on the arrival of the gate voltage pulse, has a lasting effect of reduction of the number of electrons in the steady state. This is the electronic basis of the $LT_d$. In the weak pulse intensity case, the $ST_f$ effect may be so weak that the $LT_d$ effect is predominant.

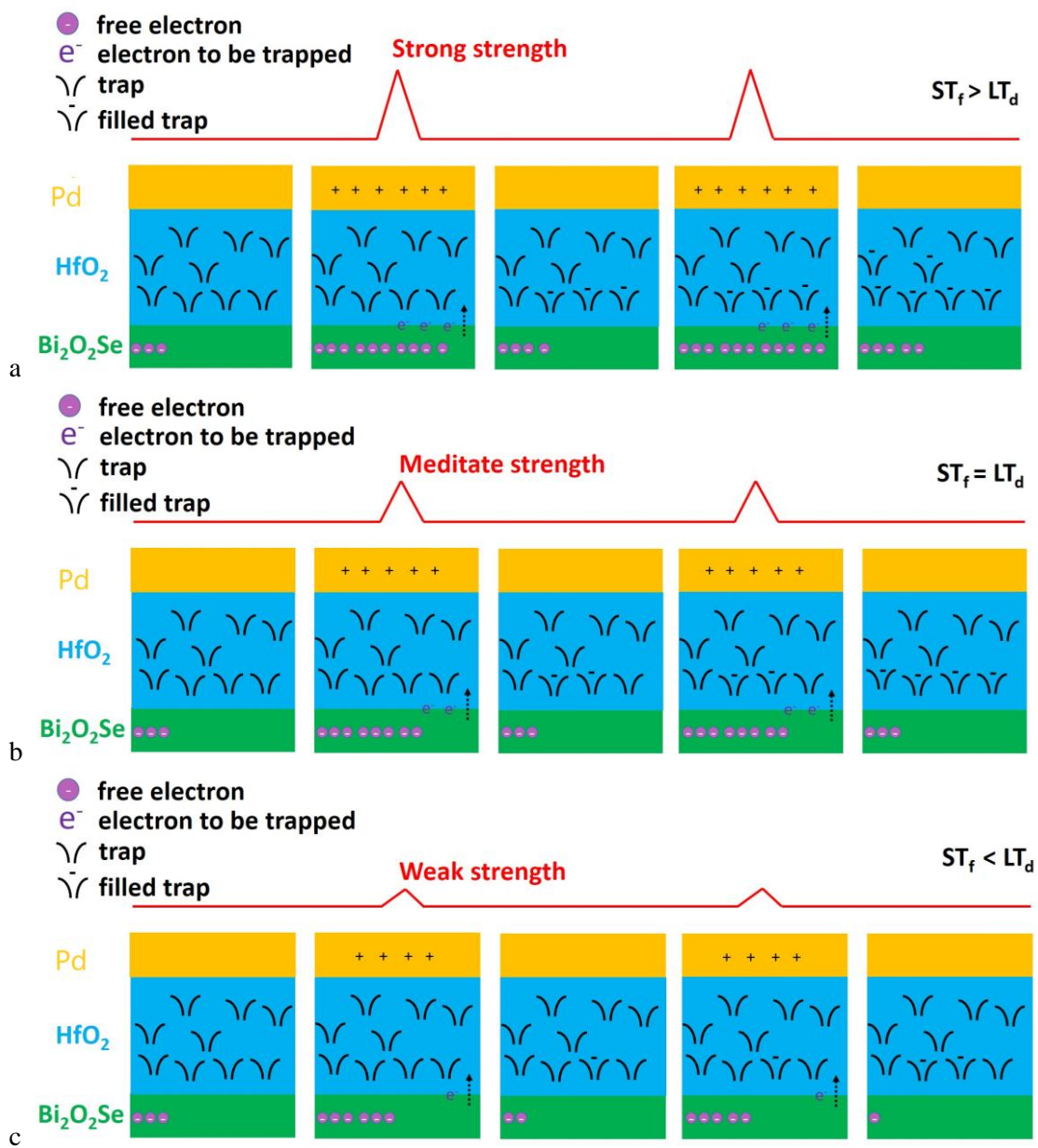

**Figure S3** Mechanisms of concomitant long-term and short-term modulation of the transmission efficacy in $Bi_2O_2Se$ synaptic device under (a) strong, (b) median and (c) weak pulse stimuli. The pulse intervals are 50 ms in all cases. Because the pulse interval is the same for all three cases, we assume that the same number of free electron carriers are reduced after the pulse elapses because of the transient gate field effect.

**S4. PSC induced using a train of pulses with variable intensities and alternating polarities.**

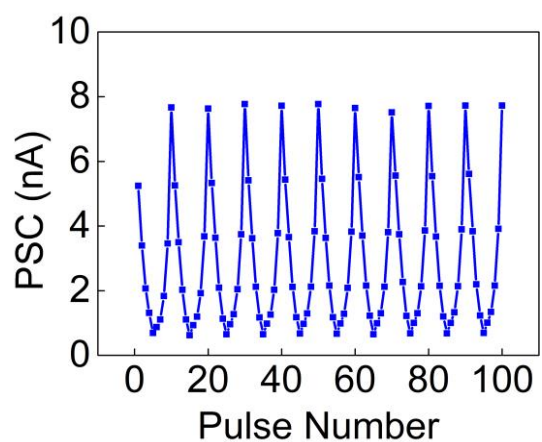

**Figure S4** PSC response to a train of alternating positive and negative $V_g$ pulses. A gradual increase in PSC is achieved using pulses with increasing pulse intensity from -1 V to -9 V in steps of -2 V. A gradual reduction in PSC is achieved using pulses with increasing pulse intensity from +2 V to +10 V in steps of +2 V. The pulse width is 50 ms and the pulse interval is 5 s.

## S5. Concomitant short- and long-term plasticity under low-voltage operation

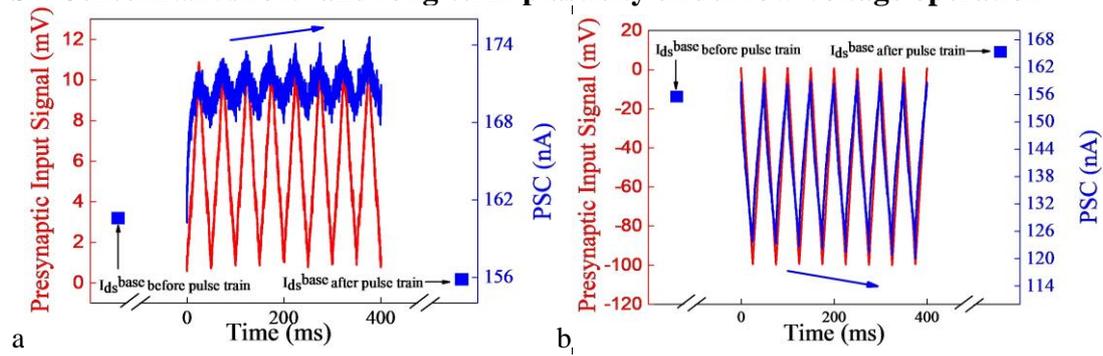

**Figure S5** (a) The concomitant $ST_f$ and $LT_d$ effects under positive pulse train measurement with low pulse intensity of 10 mV. (b) The concomitant $ST_d$ and $LT_p$ under negative pulse train measurement with low pulse intensity of 100 mV. Newly fabricated device 3 is used.

## S6. Determination of the timescale of STP

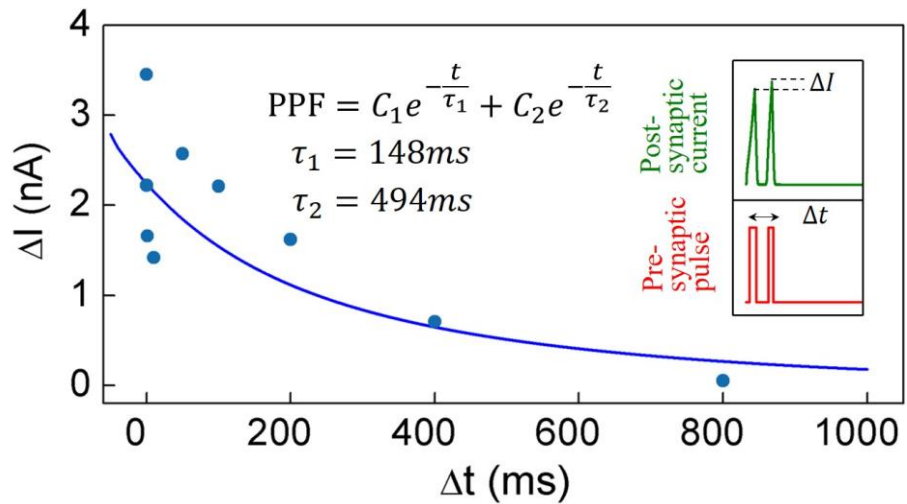

**Figure S6** Dependence of paired pulse facilitated post-synaptic current on pulse interval. Data are from pulse train measurements with pulse intensity of 6V (figure 2(b1-b3)).

In order to obtain the timescale of STP, it has to be decoupled from LTP, which is inherently difficult. A postulation is made here: the STP effect is negligible in the case of 1s pulse interval. This is from the observation of the recovery of depression effect (see **figure 2(b3)**). Based on this postulation, the STP effect can be extracted by subtracting the current values under 1s interval pulse measurement. We follow the wisdom of Burgt et al. [1] who performed data fitting to obtain the dependence of paired pulse facilitated post-synaptic current on pulse interval. In this way, a pair of time constants for our STP effect are obtained to be 148 ms and 494 ms. These values are of the same order as those of Burgt et al. who obtained approximately equivalent values to those measured in biological synapses. In addition, our time constants are below 1s, in consistence with our initial postulation.

**S7. Pulse interval-dependent working mechanisms of $Bi_2O_2Se$ synaptic device under the specific +6 V pulse intensity condition.**

In the short pulse interval case, the channel conductivity upon arrival of the next gate voltage pulse may be higher than that which occurred during the preceding pulse, and this is attributed to the presence of residual electrons. Shorter the pulse lengths cause more residual electrons to be present. This is the electronic basis of the $ST_f$. In addition to the transient effect, the possible charge trapping in the border traps, which occurs upon arrival of the gate voltage pulse, results in a lasting effect of the reduction in the number of electrons in the steady state. This is the electronic basis of the $LT_d$. In the long pulse interval case, the $ST_f$ effect may be so weak that the $LT_d$ effect is predominant.

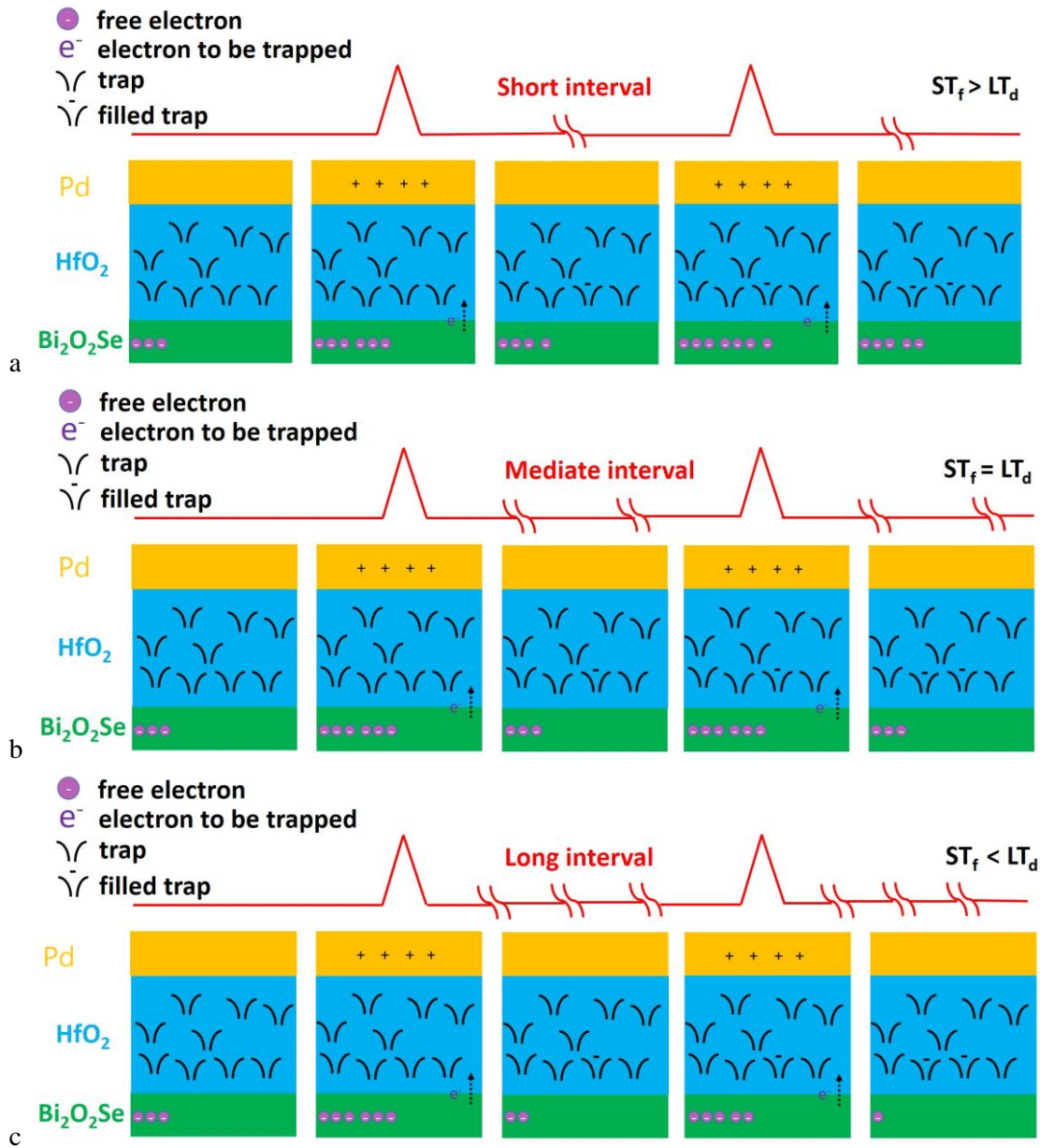

**Figure S7** Mechanisms for the concomitant long-term and short-term modulation of the transmission efficacy in $Bi_2O_2Se$ synasptic device under (a) high, (b) median and (c) low frequency pulse stimuli. The pulse intensity is +6 V in all cases. Becuase the pulse intensity is the same for all three cases, we assume that the same number of free electron carriers are increased upon pulse arrival because of the gate field effect, and that the same number of electrons are trapped.

**S8. Simulated "sleep-wake cycle autoregulation" function under the circumstances of enemy invasion during sleep.**

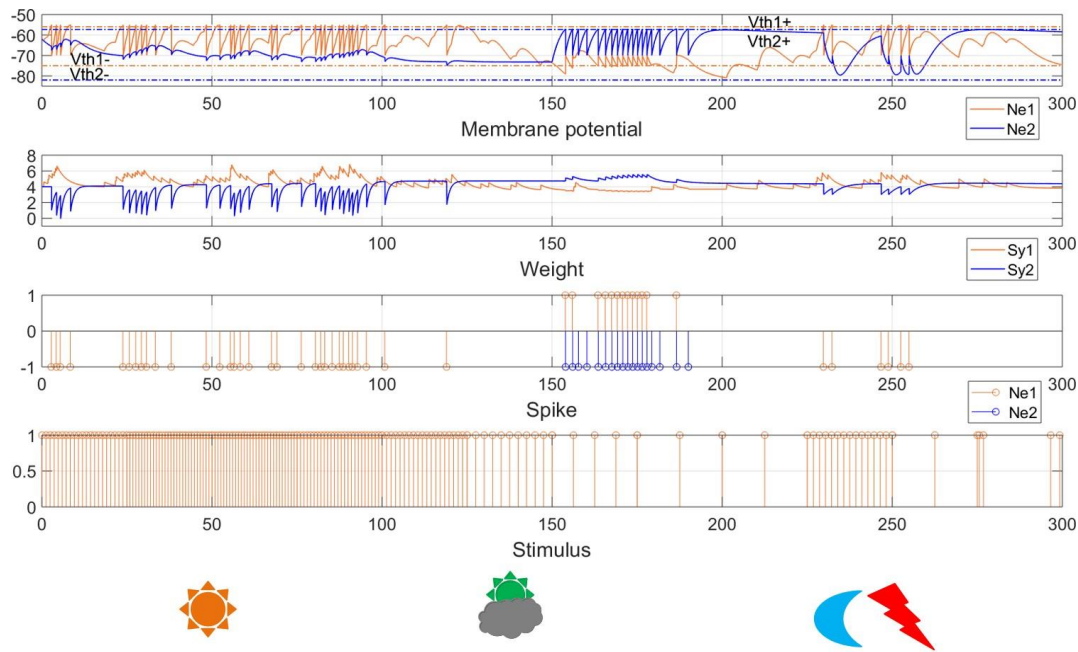

**Figure S8** Recurrent spiking neural network simulation of "sleep-wake cycle autoregulation" in the case of enemy invasion during sleep. Wake-up and locaomotion are immediately triggerred because of the $ST_f$ effect of Sy1. The changes in membrane potentials of Ne1 and Ne2, in synaptic weights of Sy1 and Sy2, and the spiking activities of Ne1 and Ne2 in response to external stimuli are presented. Enemy invasion (stimuli with higher frequency than that of the background) immediately triggers locomotion, as indicated by sudden occurrence of spiking for Ne1 at midnight.